\documentclass[10pt,a4paper,Bold]{article}

\usepackage[utf8]{inputenc}
\usepackage[T1]{fontenc}
\usepackage{graphicx}
\usepackage[font=small]{caption}
\usepackage{copyrightbox}
\usepackage{url}
\usepackage[position=top]{subfig}
\usepackage{multicol,caption}
\usepackage{mathtools, cuted}
\usepackage{lipsum, color}
\usepackage{graphicx}	
\usepackage{amsmath}	
\usepackage{amssymb}	
\usepackage{multicol}        
\usepackage{bm}		
\usepackage{pdflscape}	
\usepackage[T1]{fontenc}
\usepackage{float}
\usepackage{blindtext}
\usepackage{xcolor}
\usepackage{pgfplots}
\usepackage{times}   
\usepackage{amsmath,scalerel}
\usepackage[toc,page]{appendix}
\usepackage[sorting = none, backend = bibtex, style=numeric-comp]{biblatex}
\usepackage{abstract}

\usepackage[colorlinks=true, urlcolor=blue, linkcolor= blue,citecolor= blue, pdfborder={0 0 0}]{hyperref}
\usepackage{amsmath, nccmath}
\usepackage[none]{hyphenat}
\newenvironment{Figure}
  {\par\medskip\noindent\minipage{\linewidth}}
  {\endminipage\par\medskip}

\title{Second multicols Demo}
\author{SMH}

\usepackage{geometry}
\begin{document}
\title{A Study of thin relativistic magnetic accretion disk around a distorted black hole }
\author{Seyyed Masoud Hoseyni$^{1}$\thanks{\texttt{m.hoseyni@khayyam.ac.ir}} , 
Jamshid Ghanbari$^{1}$\thanks{\texttt{j.ghanbari@khayyam.ac.ir}} , 
Mahboobe Moeen Moghaddas$^{2}$\thanks{\texttt{Dr.moeen@kub.ac.ir}}\\
$^{1}$\ Department of Physics,Khayyam University,Mashhad,Iran\\
$^{2}$\ Department of Sciences,Kosar University of Bojnord,Bojnord, Iran\\
}
\date{\today}
\maketitle

                                                        
\begin{abstract}

 Accretion disks, swirling structures of matter spiraling into black holes, play a pivotal role in our understanding of binary star systems and their intricate evolutionary processes. While current models often simplify these complex phenomena by neglecting the influence of powerful magnetic fields, particularly within warped or distorted black hole geometries, this study delves into the crucial impact of such fields.  Focusing on a thin accretion disk encircling a Schwarzschild black hole, we meticulously investigate how the presence of a quadrupole moment, an inherent distortion in the black hole's shape, affects its spectral characteristics. By analyzing key parameters like total pressure, magnetic pressure, temperature, height scale, surface density, and radiative flux – the energy emitted by the disk – we reveal significant alterations induced by incorporating both magnetic fields and a quadrupole moment. Notably, our findings demonstrate that negative quadrupoles exert a more pronounced influence on these disk properties, highlighting the intricate interplay between these factors. This comprehensive study provides invaluable insights into the dynamics of accretion disks surrounding distorted black holes with magnetic fields, paving the way for a more accurate and nuanced understanding of these fascinating astrophysical systems. 
\\
\end{abstract}

\begin{multicols}{2}


\section{Introduction}\label{sec.Intro}

The enigmatic nature of black holes, predicted by Einstein's theory of general relativity, has spurred extensive theoretical investigation.  Early insights into the formation and behavior of relativistic accretion disks surrounding these celestial objects were provided by Page and Thorne \cite{page1974disk} and \cite{novikov1973astrophysics}, who analyzed both Schwarzschild and Kerr black holes. Further advancements came with Bardeen, Press, and Teukolsky's derivation of crucial formulae for rotating black holes \cite{bardeen1972rotating}. The pioneering work of Shakura and Sunyaev in 1973 \cite{shakura1973black} introduced the first thin accretion disk model, based on classical assumptions. This seminal work paved the way for subsequent refinements by Paczynski and Bisnovatyi in 1981 \cite{paczynski1981model},Muchotrzeb and Paczynski in 1982 \cite{muchotrzeb1982transonic}, and Abramowicz et al. \cite{abramowicz1988slim} in 1988, who delved deeper into the intricacies of thin accretion disks, refining our understanding of these fascinating astrophysical phenomena.\\
Following two decades of research, Riffer and Herold meticulously corrected an algebraic error within the structure equation of the Novikov and Thorne model \cite{riffert1995relativistic}, refining our understanding of accretion disk dynamics. Meanwhile, Doroshkevich et al. pioneered the exploration of a Schwarzschild black hole immersed in an external gravitational field with a quadrupole moment \cite{doroshkevich1965gravitational}. Their work not only constructed the metric for this complex system but also unveiled a crucial rule governing the event horizon. Geroch and Hartle delved deeper into the general properties of distorted Schwarzschild spacetime \cite{geroch1982distorted}, laying the groundwork for subsequent investigations. Subsequent studies by Frolov, Tomimatsu, and others \cite{frolov1983vacuum,tomimatsu1984distorted,frolov1986vacuum,fairhurst2001distorted,Yazadjiev:2000by,fairhurst2001isolated} focused on the specific characteristics of both static and rotating distorted black holes, solidifying their significance for understanding accretion disk behavior around these objects. Chandrasekhar further contributed to this field in 1998 by introducing equilibrium conditions for a black hole within a static external gravitational field, encompassing multipolar interactions \cite{chandrasekhar1998mathematical}.\\
Further advancements in understanding distorted black holes emerged from the work of Tomimatsu and Frolov, who presented axisymmetric static solutions describing such distortions \cite{Tomimatsu:2005td,Frolov:2007xi}. Yoshino's investigation into the formation of distorted apparent horizons provided compelling support for the hoop conjecture \cite{yoshino2008highly}. Simultaneously, Abdolrahimi and collaborators conducted extensive studies exploring the characteristics of both Schwarzschild and Kerr black holes with greater precision \cite{abdolrahimi2009interior, abdolrahimi2010distorted, abdolrahimi2014distorted, abdolrahimi2015thermodynamic, shoom2015distorted, abdolrahimi2015distorted, abdolrahimi2015properties}. Shoom and Grover delved into the complex geodesic motion of particles within the gravitational field of a distorted Schwarzschild black hole, enriching our understanding of its dynamics \cite{shoom2016geodesic,grover2018multiple}.\\ 
The exploration of magnetized black holes, a concept first introduced in 1976 by considering the nonlinearity of the Schwarzschild black hole coupled with Melvin's magnetic universe \cite{ernst1976black}, has captivated the scientific community. These objects possess significantly more complex asymptotic behavior compared to their non-magnetized counterparts \cite{aliev1989exact}. The interest in magnetized black holes stems from the crucial role their magnetic fields play in powering some of the universe's most energetic phenomena. These fields are not passive; they are instrumental in launching jets of matter emanating from the centers of galaxies and microquasars. Kunz et al. made significant strides by constructing exact solutions describing distorted black holes within external magnetic fields \cite{kunz2019distorted}. Further research has delved into the intricate interactions between Schwarzschild black holes and antisymmetric mass distributions outside their horizons, employing Einstein's field equations \cite{faraji2020thin}. The influence of these external mass distributions on relativistic thin accretion disks surrounding distorted Schwarzschild black holes with quadrupole moments has also been investigated. Layeghi et al., in a recent study \cite{layeghi2024study}, meticulously examined the alterations in the quadrupole moment for both rotating and non-rotating states of these magnetized black holes. Meanwhile, Faraji et al. have extended their research to explore relativistic thick accretion disk models incorporating both quadrupole moments and magnetic fields \cite{faraji2021magnetised}. \\
While substantial progress has been made in understanding accretion disk behavior, a critical gap remains in our comprehension of how spacetime distortion and magnetic fields interact to shape these dynamic systems. This study aims to shed light on this intricate interplay by investigating the spectral characteristics of thin accretion disks in the presence of a quadrupole moment. To achieve this objective, we proceed in a structured manner. Section \ref{sec.DM} delves into the metric and relevant equations governing distorted spacetime. Section \ref{sec.EQ} lays out the fundamental assumptions and conservation equations, considering the stress-energy tensor for a magnetized state.  In Section \ref{sec.conclusion}, we present the derived equations and visualize them through graphical representations. Finally, Section \ref{sec.discus} provides a comprehensive discussion of the obtained results, highlighting their implications for our understanding of accretion disk dynamics. Throughout this analysis, all quantities are expressed in the CGS system of units.
\\

\section{Distorted Metric}\label{sec.DM}
The Schwarzschild metric, a solution to Einstein's vacuum field equations, depicts a spherically symmetric and static black hole with asymptotic flatness and a well-defined event horizon. However, the realm of black hole solutions extends beyond this idealized case.  When considering non-isolated black holes, influenced by external mass distributions, we encounter solutions characterized by regular event horizons but lacking asymptotic flatness. This departure from isolation arises in scenarios such as binary systems where a black hole interacts gravitationally with a companion star or accretes matter into a disk. These interactions distort the surrounding spacetime, creating intricate geometries that offer fertile ground for both numerical and analytical investigations. Researchers have employed a range of techniques to explore these distorted spacetimes. Numerical simulations provide detailed insights into complex binary systems, while analytical models offer simplified yet illuminating representations of the distortions.   In our analysis, we adopt a simplifying assumption by neglecting the black hole's rotation. This approach, akin to Geroch and Hartle's pioneering work \cite{geroch1982distorted}, enables us to approximate the spacetime outside the black hole as a distorted Schwarzschild geometry under the influence of a static, axially symmetric external mass distribution.  This framework draws upon the insightful treatment presented in Chapter 10 of Griffiths' comprehensive text \cite{griffiths2009exact}.\\
The Weyl metric:
\begin{equation}
  ds^2=-e^{2U}dt^2+e^{-2U}[e^{2\gamma}(d\rho^2+dz^2)+\rho^2d\phi^2],\label{eq:1}
\end{equation}
where U and $\gamma$ are now functions of $\rho$ and z and it is usually assumed that
$t\in(-\infty,+\infty), \rho\in[0,+\infty), z\in(-\infty,+\infty), \phi\in[0,2\pi]$. \\
With $\Lambda=0$, the vacuum field equations for the metric \eqref{eq:1} imply that
\begin{equation}
  U_{,\rho\rho}+\frac{1}{\rho}U_{,\rho}+U_{,zz}=0.\label{eq:2}
\end{equation}
This can be recognised as Laplace’s equation $\nabla^{2}U=0$. Putting $\rho=rsin\theta$ and $z=rcos\theta$, the metric \eqref{eq:2} takes the form
\begin{equation}
  ds^2=-e^{2U}dt^2+e^{-2U}[e^{2\gamma}(dr^2+r^2d\theta^2)+r^2sin\theta^2 d\phi^2].\label{eq:3}
\end{equation}
It is convenient to rewrite the metric in prolate spheroidal coordinate (t,x,y,$\phi$) \cite{faraji2020thin}:
\begin{equation}
\begin{split}
  ds^2&=-\frac{x-1}{x+1} e^{2U}dt^2+M^2 (x+1)^2 e^{-2U} \\
  &\times[e^{2\gamma}(\frac{dx^2}{x^2-1} +\frac{dy^2}{1-y^2})+ (1-y^2)d\phi^2],\label{eq:4}
\end{split}
\end{equation}
where $x\in(1,+\infty), y\in[-1,+1]$ and The relation to Schwarzschild coordinates is given by $x=\frac{r}{M}-1 , y=cos\theta$. $M$ is a parameter that can be identified as the mass of the black hole. According to the \cite{layeghi2024study} we have
\begin{equation}
U=\sum_{n=0}^\infty a_n R^n P_n (\frac{xy}{R}).
\end{equation}
\begin{equation}
R=\sqrt{x^2+y^2-1}.
\end{equation}
\begin{equation}
\gamma=\sum_{n,k=1}^\infty \frac{nk}{n+k} a_n R^n P_n.
\end{equation}
The coefficients $a_n$ are known as multipole moments. These values quantify the distortion of the black hole's spacetime caused by an external field. When all multipole moments are zero the spacetime is described by the spherically symmetric Schwarzschild solution. so $n=2$ is the quadrupole moment. We adapted the notation q for the quadrupole moment, instead of $a_2$. The multipole moments must satisfy the following condition:
\begin{equation}
\sum_{n\gg 0} a_{2n+1}=0.
\end{equation}
In these coordinates, the Schwarzschild solutions $U$ and $\gamma$ read as
\begin{equation}
U=\frac{1}{2}\ln\frac{x-1}{x+1},
\end{equation}
\begin{equation}
\gamma=\frac{1}{2}\ln\frac{x^2-1}{x^2-y^2}.
\end{equation}
As \cite{layeghi2024study} mentioned: 
\begin{equation}
 B = [x+1+(x-1)ab]^2+[(1+y)a+(1-y)b]^2.
\end{equation}
\begin{equation}
a=-\alpha e^{[2q(x-y)(1+xy)]}.
\end{equation}
\begin{equation}
b=\alpha e^{[2q(x+y)(1-xy)]}.
\end{equation}
\begin{equation}
\mathcal{A}=\frac{B}{x^8}.
\end{equation}
\begin{equation}
\mathcal{B}=\frac{1}{Mx^3\Omega}.
\end{equation}
\begin{equation}
\mathcal{C}=\mathcal{B}^2 (E-\Omega L)^2.
\end{equation}
\begin{equation}
\mathcal{D}=\frac{1}{x^8}(Ae^{2U}+4a_*^2).
\end{equation}
\begin{equation}
\mathcal{E}=\mathcal{A}+3a_*^2(x^{-4}-2x^{-6}+a_*^2x^{-8}).
\end{equation}
\begin{equation}
\mathcal{F}=\frac{L\sqrt{\mathcal{C}}}{Mx}.
\end{equation}
\begin{equation}
\mathcal{G}=E\sqrt{\mathcal{C}}.
\end{equation}
\begin{equation}
\mathcal{L}=\frac{2r^2}{3M} \mathcal{B} (\mathcal{C})^{\frac{1}{2}}f.
\end{equation}

\begin{equation}
 \begin{split}
     f&=\frac{3}{2M}\frac{1}{x^2(x^3-3x+2a_*)} \\
       & (x-x_0-\frac{3}{2}a_* \ln\frac{x}{x_0}-\frac{3(x_1-a_*)^2}{x_1(x_1-x_2)(x_1-x_3)}\\
       & \ln (\frac{(x-x_1)}{(x_0-x_1)}-\frac{3(x_2-a_*)^2}{x_2(x_2-x_1)(x_2-x_3)} \ln(\frac{(x-x_2)}{(x_0-x_2)}\\
       & -\frac{3(x_3-a_*)^2}{x_3(x_3-x_1)(x_3-x_2)} \ln(\frac{(x-x_3)}{(x_0-x_3)},
  \end{split}
\end{equation}
where
\begin{equation}
x_1 = 2cos(\frac{1}{3} arccos(a_*) - \frac{\pi}{3})
\end{equation}
\begin{equation}
x_2 = 2cos(\frac{1}{3} arccos(a_*) + \frac{\pi}{3})
\end{equation}
\begin{equation}
x_3 = -2cos(\frac{1}{3} arccos(a_*))
\end{equation}
\begin{equation}
x_0=[(3+z_2-signum (a_*))\cdot \sqrt((3-z_1)(3+z_1+2\cdot z_2))]^\frac{1}{2}
\end{equation}
\begin{equation}
z_1=1+(-a_*^2+1)^\frac{1}{3}(1+a_*)^\frac{1}{3}+(1-a_*)^\frac{1}{3} 
\end{equation}
\begin{equation}
z_2 = (3 a_*^2 + z1^2)^\frac{1}{2}
\end{equation}

where $a_*=\frac{a}{M}$ and $a$ is specific angular momentum of hole. In Schwarzschild metric $a_*\simeq0$. so we derive new equations and the results will change. and $g_{\mu\nu}$ will be:

\begin{equation}
\begin{medsize}
\setlength{\arraycolsep}{-5pt}
\begin{split}
 g_{\mu\nu}&=\\ 
 &\left(\begin{array}{cccc}
-\frac{(x-1)e^{2U}}{(x+1)} & 0 & 0 & 0	\\
0 & \frac{M^2(x+1)^2 e^{2\gamma-2U}}{x^2-1} & 0 & 0	\\
0 & 0 & \frac{M^2(x+1)^2e^{2\gamma-2U}}{1-y^2} & 0	\\
0 & 0 & 0 & M^2(x+1)^2(1-y^2)e^{-2U}
\end{array}\right)
\end{split}
\end{medsize}
\end{equation}

\section{Equations}\label{sec.EQ}

Accretion disks are fundamental structures underpinning numerous astrophysical phenomena. While significant strides have been made in comprehending these intricate systems within the simplified framework of Schwarzschild spacetime, recent research reveals that many black holes deviate from this idealized model. Distortions arising from factors such as angular momentum distribution, gravitational wave interactions, or other astrophysical perturbations exert a profound influence on the dynamics of surrounding accretion disks. This inherent complexity necessitates the establishment of well-defined assumptions to guide our analytical and computational endeavors.  Only through carefully constructed models can we accurately capture the characteristics and behavior of these dynamic systems.\\
This section establishes the foundational framework for our analysis by elucidating the structure of a thin accretion disk.  We adopt the widely recognized thin disk model, as detailed in seminal works by Novikov and Thorne \cite{novikov1973astrophysics} and Shakura and Sunyaev \cite{shakura1973black}. This model provides a simplified yet conceptually robust representation of geometrically thin, optically thick, and cold accretion disks, proving suitable for our purposes. Within this framework, the disk efficiently converts a substantial fraction of its rest mass energy into radiation, with a mass accretion rate approximating the Eddington limit.  Our primary assumptions are as follows: the disk exhibits optical thickness, geometrical thinness, and remains cold; radiation emanates vertically; the disk resides in the equatorial plane; and we consider a standard relativistic thin disk surrounding a non-rotating black hole.\\
The intricate dynamics of accretion disks, the swirling vortexes of plasma encircling compact objects such as black holes and neutron stars, are profoundly shaped by magnetic fields.  These invisible forces intricately intertwine with fluid motion and gravitational influences, profoundly impacting energy transport, heating processes within the disk, and even the generation of powerful relativistic jets that shoot outward from these cosmic powerhouses. To unravel this complex interplay, we turn to a set of fundamental equations that capture the behavior of magnetic fields within these dynamic systems.
In this analysis, we posit that the magnetic field within the accretion disk originates solely from the disk's own properties and dynamics, independent of any influence exerted by the central black hole's magnetic field. This assumption allows us to isolate the effects of the disk itself on the system under investigation. \\
Relativistic thin accretion disk models aim to elucidate the intricate dynamics of matter accreting onto compact astrophysical objects like black holes or neutron stars. These models employ a set of three fundamental equations to meticulously describe the radial structure of the disk and the motion of the accreting material. 
The first equation is the particle number conservation \cite{faraji2020thin}:
\begin{equation}
(\rho u^\mu )_{;\mu}=0.\label{eq:21}
\end{equation}
In which $u^\mu$ is the four velocity of the fluid and $\rho$ is the rest mass density.\\
The Second equation is the radial momentum equation, reads as:
\begin{equation}
h_{\mu\sigma}(T^{\sigma\nu})_{;\nu}=0,\label{eq:22}
\end{equation}
where $h^{\mu\nu}=u^\mu u^\nu + g^{\mu\nu}$ (the projecion tensor), and\\
$T^{\sigma\nu}$ is the stress-energy tensor.\\
The third one is the conservation of relativistic induction which is given as \cite{faraji2023magnetized}:
\begin{equation}
(b^{\mu}u^{\nu}-b^{\nu}u^{\mu})_{;\nu}=0,\label{eq:23}
\end{equation}
Where $b_{\mu}$ is the magnetic field four-vector, which reduces to the ordinary magnetic field in the fluid frame.\\
The stress-energy tensor is expressed below:
\begin{equation}
T^{\mu\nu}=\varepsilon u^{\mu}u^{\nu}+(p+\frac{b^2}{2})g^{\mu\nu}-b^{\mu}b^{\nu},
\end{equation}
\begin{equation}
\varepsilon=\rho+u+p+b^2,
\end{equation}
\begin{equation}
b^2=b^{\mu}b_{\mu},
\end{equation}
Where $\varepsilon$ is the enthalpy and P is the total pressure. The quantity  $b_{\mu}$ represents the components of the magnetic field. In the fluid's rest frame, its magnitude squared is expressed as: 
\begin{equation}
b^2 = 2 \mu_0 P_m
\end{equation}
where $P_m$ denotes the magnetic pressure.\\
By placing metric components, we have obtained from distorted metric and determined the total energy, angular velocity and angular momentum follows \cite{faraji2020thin}.
The angular momentum:
\begin{equation}
\Omega=\frac{-(g_{t\phi})_{,r}+\sqrt{(-(g_{t\phi})_{,r})^2-(g_{\phi\phi})_{,r}(g_{tt})_{,r}}}{g_({\phi \phi})_{,r}}.
\end{equation}
The total energy:
\begin{equation}
E=\frac{g_{tt}+\Omega g_{t\phi}}{\sqrt{-g_{tt}-2\Omega g_{t\phi}-\Omega^2 g_{\phi\phi}}}.
\end{equation}
The angular velocity:
\begin{equation}
L=\frac{\Omega g_{\phi\phi}+g_{t\phi}}{\sqrt{-g_{tt}-2\Omega g_{t\phi}-\Omega^2 g_{\phi\phi}}}.
\end{equation}
so
\begin{equation}
\Omega=\frac{e^{-q(x^2-1)}}{M\sqrt{(x+1)^3}} \sqrt{\frac{-qx^3+qx+1}{qx^2+qx+1}}
\end{equation}
\begin{equation}
E=\frac{(x-1)e^{-\frac{q}{2}(x^2-1)}}{\sqrt{x+1}} \sqrt{\frac{qx^2+qx+1}{2x^3q+x(1-2q)-2}}
\end{equation}
\begin{equation}
L=M(x+1)e^{\frac{q}{2}(x^2-1)} \sqrt{\frac{-qx^3+qx+1}{2x^3q+x(1-2q)-2}}
\end{equation}

\section{Conclusion}\label{sec.conclusion}
Through meticulous manipulation and simplification of the metric equations, we derived novel algebraic expressions that illuminate the spacetime structure of both undistorted and distorted Schwarzschild black holes. These insights are visually represented in the accompanying figures, which juxtapose the distinct geometries of these two scenarios. It is crucial to note that the distorted solutions hold validity only within a localized region proximate to the event horizon. Furthermore, the accuracy of our exact solution exhibits sensitivity to the chosen quadrupole moment, highlighting the importance of this parameter in shaping the resulting spacetime distortions. The specific values employed for these parameters are: 
\begin{equation}
\alpha=0.1
\end{equation}
\begin{equation}
M=10^7 M_{\odot}\simeq 1.99\cdot 10^{40} g
\end{equation}
\begin{equation}
L_{Edd}= 1.2\cdot 10^{38}\cdot \frac{M}{M_{\odot}} \frac{erg}{s}.
\end{equation}
\begin{equation}
\mu_0=1.3\times10^{-5} \frac{dyne}{cm^2}.
\end{equation}
\begin{equation}
c=3\cdot 10^{10} \frac{cm}{s}
\end{equation}
\begin{equation}
\dot{M}=10^{-4} \frac{M_{\odot}}{yr} .
\end{equation}
\begin{equation}
G=6.67 \cdot 10^{-8} \frac{cm^3}{g\cdot s^2}
\end{equation}
\begin{equation}
k=1.38\cdot 10^{-16} \frac{erg}{K}
\end{equation}
\begin{equation}
m_{p}=1.67\cdot 10^{-24} g
\end{equation}
Novikov and Thorne  highlights a complex structure within \cite{novikov1973astrophysics}, comprised of three distinct regions. The innermost region is characterized by the dominance of radiation pressure over gas pressure, resulting in an opacity primarily attributed to electron scattering. To accurately model the radial structure of this innermost zone, we will employ fully relativistic equations, allowing us to write the following. The Flux equation is:\\
\begin{equation}
F=7\times 10^{26} \dot M  M^{-1}  (r_*)^{-3} \mathcal{B}^{-1} \mathcal{C}^{-\frac{1}{2}} \mathcal{L}.
\end{equation}
And the Surface density:
\begin{equation}
\Sigma=5\times \alpha^{-1} \dot{M}^{-1} ({r_*})^{\frac{3}{2}} \mathcal{B}^3 \mathcal{E} \mathcal{C}^{\frac{1}{2}} \mathcal{A}^{-2} \mathcal{L}^{-1}.
\end{equation}
Height scale:
\begin{equation}
h=10^5 \dot{M} \mathcal{A}^2 \mathcal{C}^{\frac{1}{2}} \mathcal{L} ({r_*})^{-1} \mathcal{B}^{-3} \mathcal{D}^{-1} \mathcal{E}^{-1}.
\end{equation}
Temperature: 
\begin{equation}
T=2\times10^7 \mathcal{B}^{\frac{1}{2}} \mathcal{E}^{\frac{1}{4}} \alpha^{-\frac{1}{4}} M^{-\frac{1}{4}} ({r_*})^{-\frac{3}{8}} \mathcal{A}^{-\frac{1}{2}}.
\end{equation}
The pressure $(P_{gas}+P_{radiation})$: 
\begin{equation}
P_{g\&r}=5\times10^{-5} M^{\frac{7}{4}} \alpha^{-\frac{1}{4}} \dot{M}^{-2} ({r_*})^{\frac{21}{8}} \mathcal{B}^{\frac{9}{2}} \mathcal{D} \mathcal{E}^{\frac{5}{4}} \mathcal{A}^{-\frac{5}{2}} \mathcal{L}^{-2}.
\end{equation}
Magnetic Pressure:
\begin{equation}
P_B=\frac{7\times10^7}{2\mu_0} 7\times10^7 \mathcal{B} \mathcal{E}^\frac{1}{2}{r_*}^\frac{3}{4}\mathcal{A}
\end{equation}
Total pressure:
\begin{equation}
P_{total}=P_B+P_{g\&r}
\end{equation}
Magnetic field:
\begin{equation}
b=7\times10^7 \mathcal{B} \mathcal{E}^\frac{1}{2}{r_*}^\frac{3}{4}\mathcal{A},
\end{equation}
where $r_*\simeq4$. In this section, we will visually explore the impact of quadrupole moments on spacetime geometry by plotting graphs derived from equations 43 to 50.  The parameter $q$ controls the strength of the quadrupole distortion; a value of $q = 0$ represents an undistorted Schwarzschild black hole, while $q=\pm0.00001$ introduces distinct distortions. This analysis aligns with previous work by Shoom \cite{shoom2016geodesic} who investigated the effects and ranges of various quadrupole moments. To facilitate a focused examination, we have selected a specific range for the radial coordinate $x$, spanning from 3 to 15, where noticeable changes occur.\\ 
The first graph will depict the shear force as a function of $x$. For the undistorted case $q = 0$, the slope initially increases rapidly after the event horizon and then gradually decreases as we move further away.  
Crucially, in the distorted scenarios, this decreasing trend exhibits distinct behavior depending on the sign of $q$. A positive $q$ value results in a less pronounced decrease, while a negative $q$ value amplifies it. This variation underscores the influence of the quadrupole moment on the accretion disk's dynamics and flux. Notably, both graphs are situated approximately equidistant from $q=0$.\\

\begin{Figure}
 \centering
 \advance\leftskip-2cm
 \advance\rightskip-2cm
 \includegraphics[width=8cm, height=7cm]{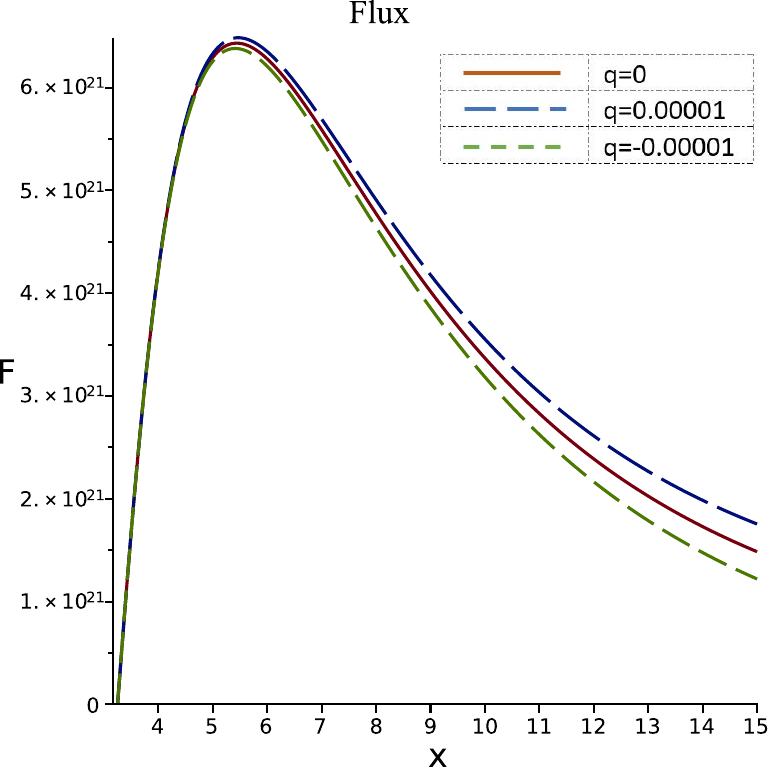}
 \captionof{figure}{Radiation flux F in the $(\frac{erg}{cm^2\cdot s})$ unit.}
\end{Figure}

The surface density plot reveals a compelling interplay between the quadrupole moment and the accretion disk's structure. A negative $q$ value induces a steeper growth in surface density as we move away from the minimum point. This pronounced effect underscores the significant influence of the quadrupole on shaping the distribution of matter within the accretion disk.  However, despite this initial steep increase, the trend eventually softens as we move further away from the event horizon, suggesting a more nuanced interplay between the quadrupole and other physical factors at larger radial distances. 
\\
\begin{Figure}
 \centering
 \advance\leftskip-2cm
 \advance\rightskip-2cm
 \includegraphics[width=8cm, height=7cm]{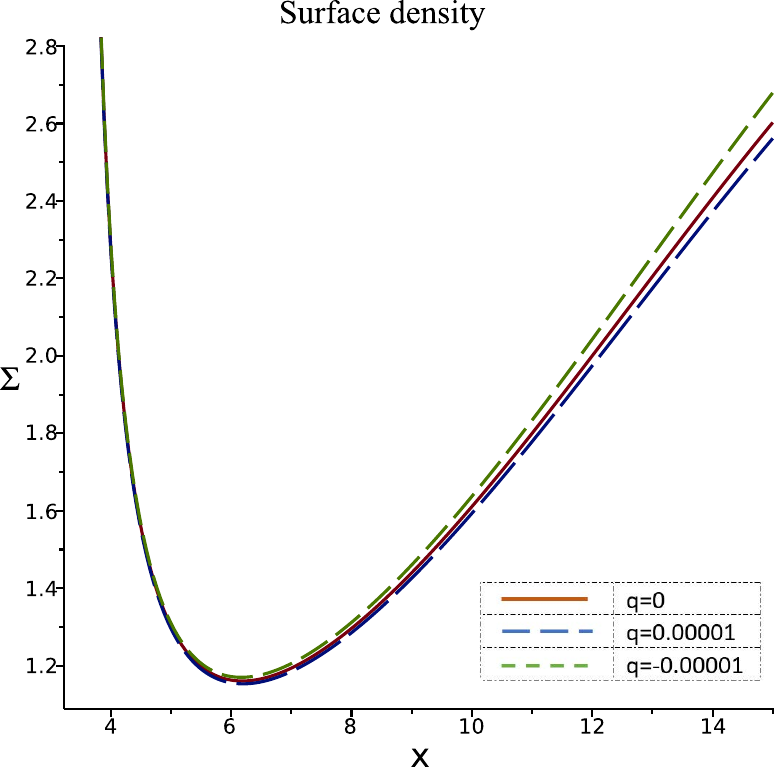}
 \captionof{figure}{Surface densiry ${\Sigma}$ in the $(\frac{g}{cm^2})$ unit.}
\end{Figure}

Examining the variations in height scale across these plots reveals a pattern consistent with previous visualizations: positive and negative $q$ values diverge symmetrically from the reference point of $q = 0$.  However,  the negative state exhibits a greater disparity from the positive state compared to their respective distances from $q = 0$. This graph (denoted as $"h"$) effectively visualizes the distinct alterations induced by the quadrupole moment on the spatial configuration.
\\
\begin{Figure}
 \centering
 \advance\leftskip-2cm
 \advance\rightskip-2cm
 \includegraphics[width=8cm, height=7cm]{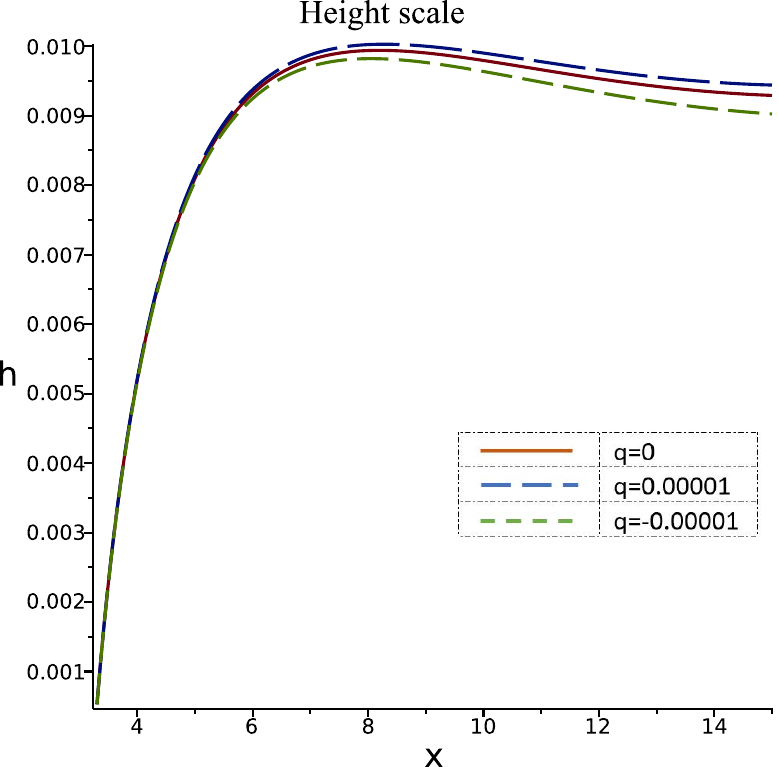}
 \captionof{figure}{Height scale h of the disk for \(q=0\) and \(q\neq0\).}
\end{Figure}

The temperature variation graph aligns with expectations, depicting a consistent decrease in temperature as the distance from the event horizon expands.  Initially, this temperature decline manifests as a uniformly sloping gradient. However, as the radial distance increases, the rate of temperature decrease intensifies, exhibiting a clear dependency on the magnitude of the quadrupole value.
\\
\begin{Figure}
 \centering
 \advance\leftskip-2cm
 \advance\rightskip-2cm
 \includegraphics[width=8cm, height=7cm]{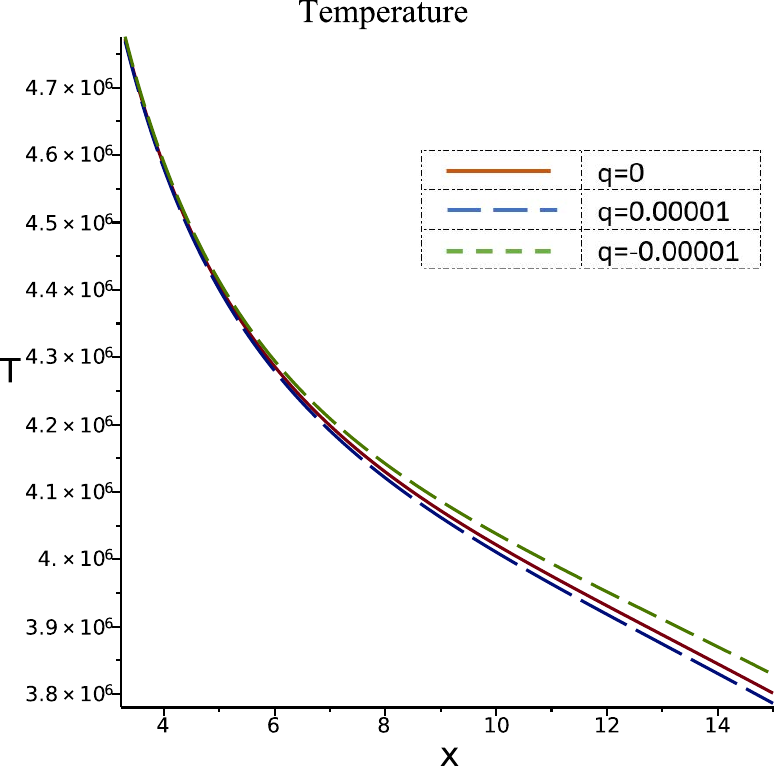}
 \captionof{figure}{Temperature in the unit k.}
\end{Figure}

The graph illustrating pressure variation against $x$ (representing the combined effect of gas pressure and radiation pressure) vividly reveals the influence of a negative quadrupole moment $q$. The alterations in pressure induced by a negative $q$ are readily apparent and substantial. Conversely, the variations attributable to positive $q$ and zero $q$ remain relatively minor and less pronounced.

\begin{Figure}
 \centering
 \advance\leftskip-2cm
 \advance\rightskip-2cm
 \includegraphics[width=8cm, height=7cm]{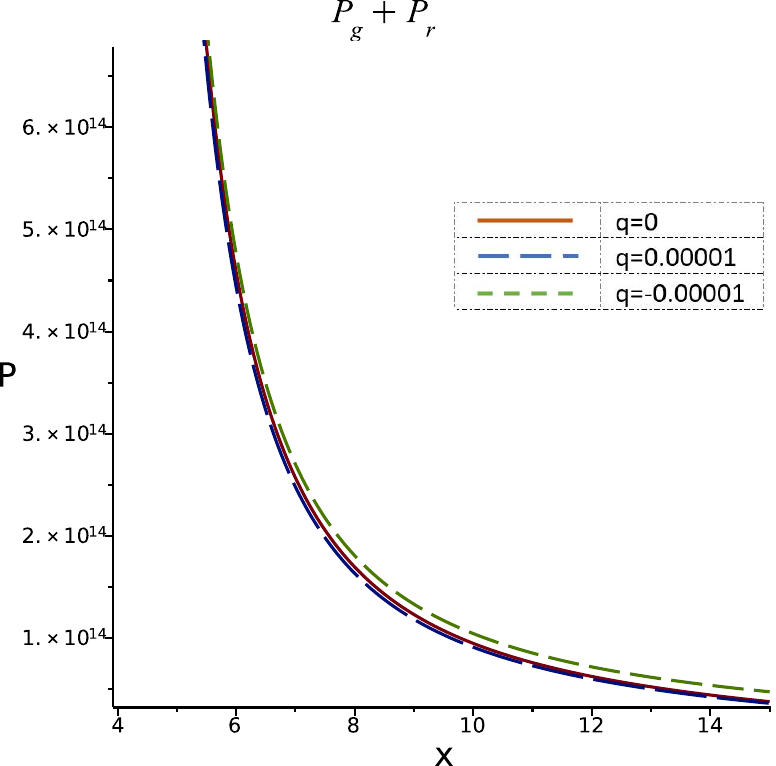}
 \captionof{figure}{Radiation pressure with gas pressure in the $(\frac{dyn}{cm^2})$ unit.}
\end{Figure}

A comparative analysis of  $P_{gr}$ and $P_B$ reveals, yet again, that variations induced by a negative quadrupole moment $q$ are more pronounced than those caused by a positive $q$. Furthermore, contrasting the total pressure with $P_{gr}$ unequivocally demonstrates the significant influence exerted by magnetic forces on the overall pressure distribution. 

\begin{Figure}
 \centering
 \advance\leftskip-2cm
 \advance\rightskip-2cm
 \includegraphics[width=8cm, height=7cm]{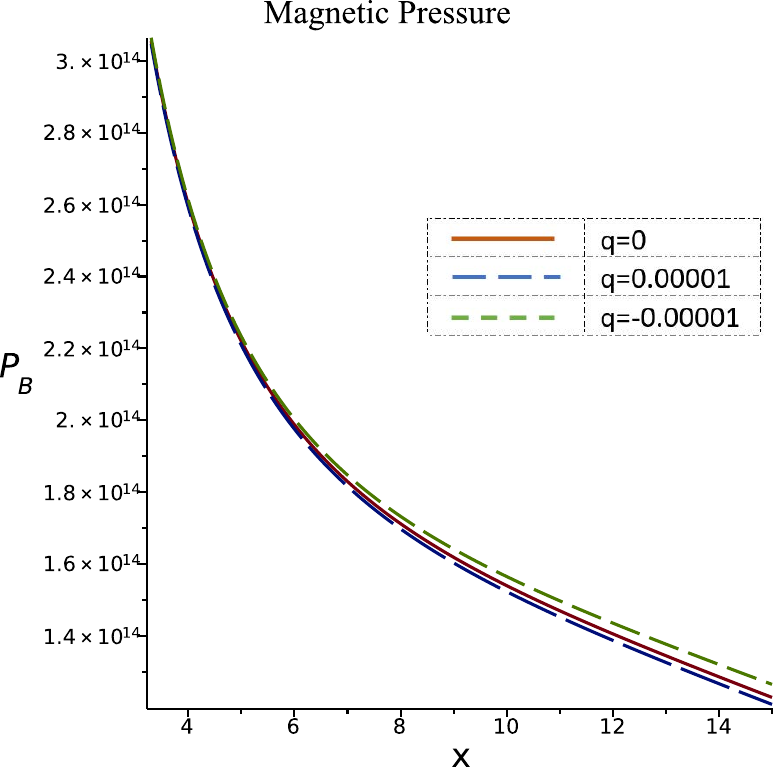}
 \captionof{figure}{Total pressure in the $(\frac{dyn}{cm^2})$ unit.}
\end{Figure}

\begin{Figure}
 \centering
 \advance\leftskip-2cm
 \advance\rightskip-2cm
 \includegraphics[width=8cm, height=7cm]{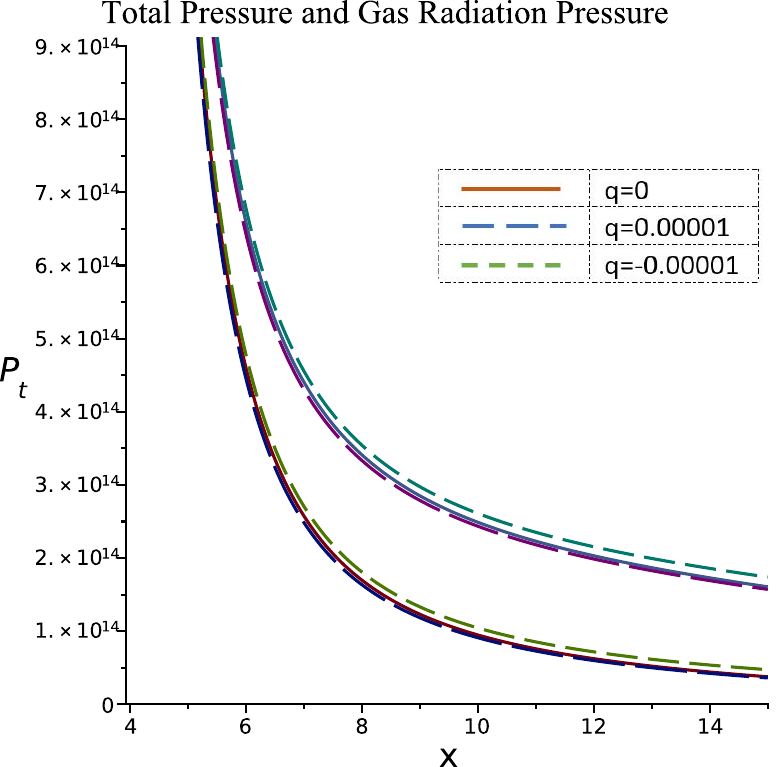}
 \captionof{figure}{Pressure in the $(\frac{dyn}{cm^2})$ unit.}
\end{Figure}

A graphical analysis of magnetic field strength plotted against distance elucidates a compelling trend: variations attributable to a negative quadrupole moment $q$ demonstrably surpass those induced by a positive $q$. As we traverse outward from the event horizon, the magnetic field strength undergoes a gradual decline, culminating in a peak value proximate to the horizon itself. 
\\
\begin{Figure}
 \centering
 \advance\leftskip-2cm
 \advance\rightskip-2cm
 \includegraphics[width=8cm, height=7cm]{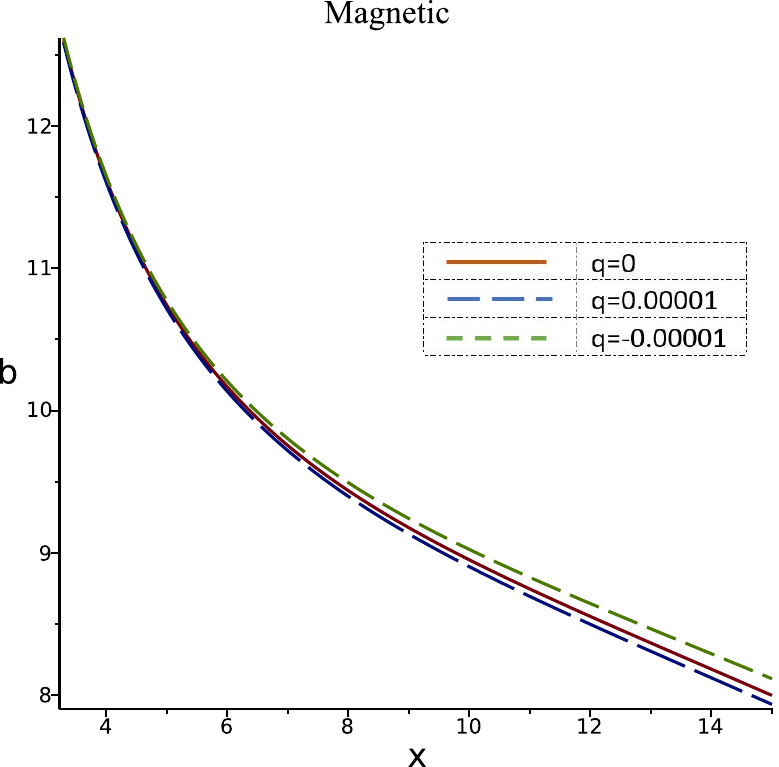}
 \captionof{figure}{Magnetic field in the gauss(G) unit.}
\end{Figure}

\section{Discussion}\label{sec.discus}
In this research, we investigated several key quantities, including force (F), surface density ($\Sigma$), temperature (T), pressure (P), and magnetic field (b), using the distorted Schwarzschild black hole metric under the influence of a magnetic field. We considered three distinct values of the quadrupole moment $q$ to demonstrate how the magnetic factor influences the distortion of the black hole. The resulting graphical data clearly show that variations linked to negative $q$ have a substantial impact on these physical parameters.\\
Recognizing the significance of this study in exploring relativistic distortions of black holes and the integration of the magnetic field, we strongly encourage future investigations to directly compare the graphs derived from negative $q$ with observational data collected from actual telescopes. Preliminary comparisons with representative accretion disk models indicate a minimal difference between our theoretical predictions and empirical observations, implying that this model may offer a more precise and realistic representation of these intricate astronomical phenomena.\\
Through this investigation, which analyzes the effects of the magnetic field and presents solutions for thin distorted disks surrounding non-rotating black holes, we lay a promising foundation for a deeper understanding of these astrophysical phenomena.

\newpage
\printbibliography

\end{multicols}




\end{document}